\documentclass[aps,twocolumn,showpacs,superscriptaddress,citeautoscript,pra,floatfix]{revtex4-1}

\usepackage[utf8]{inputenc}

\usepackage{bm}
\usepackage{graphicx,float}
\usepackage{subcaption}
\usepackage{amsmath,amssymb}
\usepackage{amsfonts}
\usepackage{color}
\usepackage{physics}
\usepackage{ragged2e}

\usepackage[colorlinks=true,%
            linkcolor=blue,%
            urlcolor=blue,%
            citecolor=blue,%
            filecolor=blue,% 
            bookmarksopen=true,%
            pdfauthor={DZ_JPRA_PAO},%
            pdftitle={MBS_BIC_MQDS},%
            pdfsubject={MBS_BIC_MQDS_Manuscript},%
            pdfpagemode=UseOutlines]{hyperref}
        
\definecolor{Blue}{rgb}{0.3,0.3,0.9}
\definecolor{Red}{rgb}{0.9,0.3,0.3}
\definecolor{Green}{rgb}{0.3,0.6,0.3}
\definecolor{Black}{rgb}{0.0,0.0,0.0}

\begin{document} % PR(B,L)
%\title{Bound states in the continuum in crossbar junctions in one-dimensional waveguides}
\title{Beyond symmetry-protected BICs: transmission through asymmetric crossbar junctions in one-dimensional waveguides}
\author{Sofia Pinto}
\affiliation{Departamento de F\'{\i}sica, Universidad T\'{e}cnica Federico Santa Mar\'{\i}a, Casilla 110 V, Valpara\'{\i}so, Chile}
%\email{sofia.pinto@sansano.usm.cl}
\author{Rafael A. Molina}
\affiliation{Instituto de Estructura de la Materia IEM-CSIC, Serrano 123, Madrid, E-28006, Spain}
%\email{pedro.orellana@usm.cl}
\author{Pedro A. Orellana}
\affiliation{Departamento de F\'{\i}sica, Universidad T\'{e}cnica Federico Santa Mar\'{\i}a, Casilla 110 V, Valpara\'{\i}so, Chile}
%\email{pedro.orellana@usm.cl}
\date{November 2023}

\begin{abstract}
Over the last few decades, the study of Bound States in the Continuum, their formation, and properties has attracted lots
%lots 
of attention, especially in optics and photonics. It is particularly noticeable that most of these investigations base their studies on symmetric systems. In this article, we study the formation of bound states in the continuum in electronic and photonic transport systems consisting of crossbar junctions formed by one-dimensional waveguides, considering asymmetric junctions with commensurable lengths for the upper and lower arms. 
%of the latter
We also study how BICs form in linear junction arrays as a function of the distance between consecutive junctions
%the effect of forming these BICs caused by the distance variation between consecutive junctions 
and their commensurability with the upper and lower arms. We solve the Helmholtz equation %in the presence of 
for
the crossbar junctions and calculate the transmission probability, probability density in the intersections, and quality factor. The presence of quasi-BICs is reflected in the transmission probability as a sharp resonance in the middle of a symmetric Fano resonance along with Dirac's delta functions in the probability density and divergence in the quality factors.
\end{abstract}

\maketitle

\section{Introduction} 

Bound states in the continuum (BICs) are states that remain spatially localized and with no decay despite coexisting with the continuum of radiation spectrum of the system \cite{Hsu16,Sadreev2021}. Von Neumann and Wigner \cite{vonNeumannWigner29} first predicted them for quantum mechanics. In their work, they found a solution for the Schrödinger equation with a particular oscillating potential in which a bound state with discrete and positive energy coupled to the radiation continuum was formed due to multiple interference processes, resulting in the complete suppression of energy leakage.
%\cite{koshelev2020engineering}. 
%Although the particular model proposed by these scientists has not been implemented experimentally, the discovery has laid the foundation for the development of the investigation of these states in various areas of physics, such as acoustics and optics.\\
%After this first discovery, the study of this subject lost strength. 
For many decades, the results of Von Neumann and Wigner were considered nothing but a mathematical curiosity, probably due to the artificial properties of the potential they used to show the properties of BICs. In the 1970s and subsequent years, various theoretical studies presented energy states with the characteristics already mentioned, but %these 
the authors of those works did not relate the results obtained to the phenomenon of BICs\cite{koshelev2020engineering}. Then, in 1985, Friedrich and Wintgen reformulated the concept of the BIC of Von Neumann and Wigner in a more general framework as the result of complete destructive interference of two resonances undergoing an avoided crossing \cite{Friedrich1985}. More recent works showed that BICs could be made robust through symmetry
arguments \cite{Bulgakov2006,Moiseyev2009}. In this case, the coupling to the continuum is forbidden due to the conservation of some symmetry. BICs were, then, divided into two categories: the accidental ones, like in the former works by Friedrich and Wintgen, and the symmetry-protected ones \cite{Hsu16}. A new category of BICs can be related to large degeneracies induced by general lattice symmetries \cite{FernandezHurtado2014,Mur14,Mur2020}. An important development was the realization that, on many occasions, a small symmetry breaking in the case of symmetry-protected BICs or a small change of parameters in the case of accidental BICs induce the appearance of Fano-like resonances that were later named quasi-BICs \cite{Guevara2003,Guevara2006,Koshelev2018}. 

BICs, although first theoretically predicted in quantum mechanics, are a general wave phenomenon, which is reflected in the experimental situation.
%Later, in 1992, Capasso, Sirtori, Faist et al. 
The first reported evidence for the formation of these states was made in an electronic system, a semiconductor heterostructure \cite{capasso1992observation}.  %but 
However, it wasn't until 2008
%, when Marinica et. al. performed studies concerning the physics of these states in optics, 
that symmetry-protected BICs were measured for the first time in an optical waveguide array \cite{Marinica08}.
In the last decade, BICs' investigation has become an active topic with experiments in photonics, phononics, plasmonics, and others,  mainly due to its device manufacturing applications \cite{shi2022planar}. With this increase of attention came an increase in the number of studies on different systems that may hold these exotic states \cite{plotnik2011experimental,Vicencio15,huang2021sound}, most of which base their research on symmetric properties. Accidental BICs have also been measured \cite{Hsu13,Koshelev2018} and may have important applications as high-Q narrow frequency resonators \cite{Jin2019}. Other applications of BICs include sensors \cite{Abujetas2019,Quotane2022,Jacobsen2022}, lasers \cite{Kodigala2017,Ha2018}, filters \cite{Foley2014}, transducers \cite{Yu2020} and actuators \cite{Qin2022}.

In this work, we go beyond the symmetry-protected BICs paradigm and show that BICs can also appear in asymmetric crossbar structures when their arms show the commensurability of their dimensions. Our analysis relies on a simple scattering formalism of single-channel waveguides, which allows us to find analytical solutions and a deep understanding of the BIC formation mechanisms. The signatures of BICs in the transmission show sharp resonances in the middle of antiresonances when the commensurability between the lengths of the side-attached structures breaks slightly (quasi-BICs as they are called in the literature \cite{Cortes2014,Xu2019,Liu2019}). We also extend our analysis to two or more crossbar structures in series. New BICs form due to hybridizing the wave functions in the side-attached bars and the central channel. We also study the formation of bands for many scattering units. Finally, we discuss the possibility of using our setup as a sensor for the impurities or imperfections of the systems.

    \begin{figure}[H]
        \centering 
        \includegraphics[width=1\columnwidth]{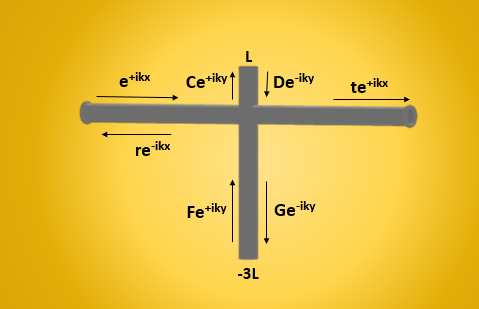}
        \includegraphics[width=1\columnwidth]{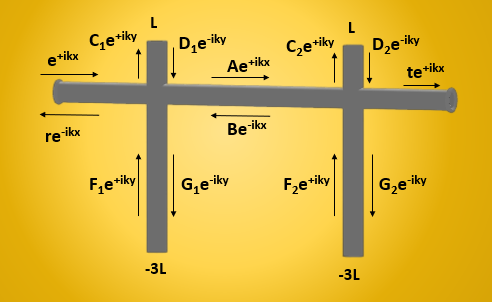}%\includegraphics[width=\textwidth]{Fig1.pdf}

        \caption{Two setups consisting of a single crossbar junction in a one-dimensional waveguide (top) and a double crossbar junction (bottom). The system is open through its horizontal arms and remains closed at the ends of its vertical arms, being their respective lengths $L^+$ (in the cases shown $L$) and $L^-$ (in the cases shown $3L$).}
        \label{Fig1}
    \end{figure}

%\section{Studied Models and Results}
\section{Models and Results}
\subsection{Single Crossbar Junction}
%{\color{red}First, we study the simplest case, a system consisting of a single crossbar junction as shown in Figure \ref{Fig1}, considering particles approaching from the left side of the latter. The setup composes a finite crossbar structure with upper and lower arms, with lengths $L^+$ (upper) and $L^-$ (lower), respectively. }

First, we study the simplest setup, a system consisting of a single crossbar junction of single channel waveguides as shown in Figure \ref{Fig1}. The incident waves approach from the left and can be transmitted or reflected at the junction. The upper and lower arms are finite with lengths $L^+$ (upper) and $L^-$ (lower), respectively. Note that this setup can be considered a variation of a Michelson-Morley interferometer \cite{Born1964}.
However, the differences are important and allow bound states in vertical arms. Under particular circumstances, some of these bound states can decouple from the continuum in the horizontal waveguide. When these BICs form, there are clear signatures in the transmission spectrum that we analyze below.

%We will first study the simplest case, a system that consists on a single crossbar junction as shown in Figure \ref{Fig1}, considering particles approaching from the left side of the latter. The system is open through its horizontal arms and remains close at the ends of its vertical arms, being their respective lengths $L^+$ (upper) and $L^-$ (lower).

In Appendix \ref{AppA}, we explain the analytic calculations for the scattering problem in this system. The final result for the total transmission probability is:

    \begin{equation} \label{15}
    T=\abs{t}^2= \frac{4}{4+[cot(kL^+)+cot(kL^-)]^2}
    \end{equation}

To understand the result, we apply the following variable changes
    \begin{equation} \label{39}
        L^+=n+\frac{\Delta}{2}
    \end{equation}
    \begin{equation} \label{40}
        L^-=m-\frac{\Delta}{2}
    \end{equation}
%    \begin{equation} \label{41}
%        a=l\pm \nu
%    \end{equation}
The variables "n" and "m" are integers that are measured in some unit of length. By making these changes, it becomes easier to analyze the impact of having the top and bottom sidearm lengths that are commensurate. We simplify the final expressions by renaming $k / n$ as $k'$.

%We'll also work with units of  $\frac{k}{\pi}$ instead of $k$ to simplify the analysis. 
%\subsection{Single Crossbar Junction}
We then rewrite Eq. (\ref{15}) in the new variables:  
%using changes \ref{39} to \ref{41}:
    \begin{equation} \label{42}
    T= \frac{4sin^2(\pi k'(n+\frac{\Delta}{2}))sin^2(\pi k'(m-\frac{\Delta}{2}))}{4sin^2(\pi k'(n+\frac{\Delta}{2}))sin^2(\pi k'(m-\frac{\Delta}{2}))+sin^2(\pi k'(n+m))}
    \end{equation}
Analyzing the previous expression we see that, for $\Delta=0$, there will be symmetric Fano resonances \cite{limonov2021fano} for every value of $k'$ that meets the following conditions:
    \begin{equation} \label{43} 
        k'=\frac{s}{n} \hspace{0.5cm} \bm{and/or} \hspace{0.5cm} k'=\frac{p}{m}  \hspace{0.5cm}  \forall (s,p)\in \mathbb N
    \end{equation}

%%%%% HE CAMBIADO Z por N porque momentos negativos solamente cambian el sentido del problema de derecha 
%%%%% a izquierda
Beside this, the value of \emph{T} will be undetermined for $\Delta=0$ and every value of $k'$ that meets:
    \begin{equation} \label{44}
       \exists(s,p)\in \mathbb N \hspace{0.5cm}:  \hspace{0.5cm} 
       k'=\frac{s}{n}=\frac{p}{m} \hspace{0.5cm}
    \end{equation}
This mathematical condition of commensurable lengths marks the presence of a BIC in the system at these values of the renormalized momentum $k'$. As BICs do not couple to the continuum, they cannot be seen in the transmittance of the system. However, any small symmetry break turns the BIC into a quasi-BIC, which appears as a narrow spectrum resonance. So for $\Delta \rightarrow 0$ and the same commensurability condition, we obtain $\emph{T}=1$. This is the experimental signature of BICs in the transmission spectrum.

%(a BIC) or equal to 1 for $\Delta\approx0$ (a quasi-BIC), for every value of $k$ that meets:
%    \begin{equation} \label{44}
%       \exists(s,p)\in \mathbb N \hspace{0.5cm}:  \hspace{0.5cm} k=\frac{s}{n}=\frac{p}{m} 
%    \end{equation}

\begin{figure}[H]
\centering
%    \begin{subfigure}
%    {0.5\textwidth}
    \includegraphics[width=\columnwidth]{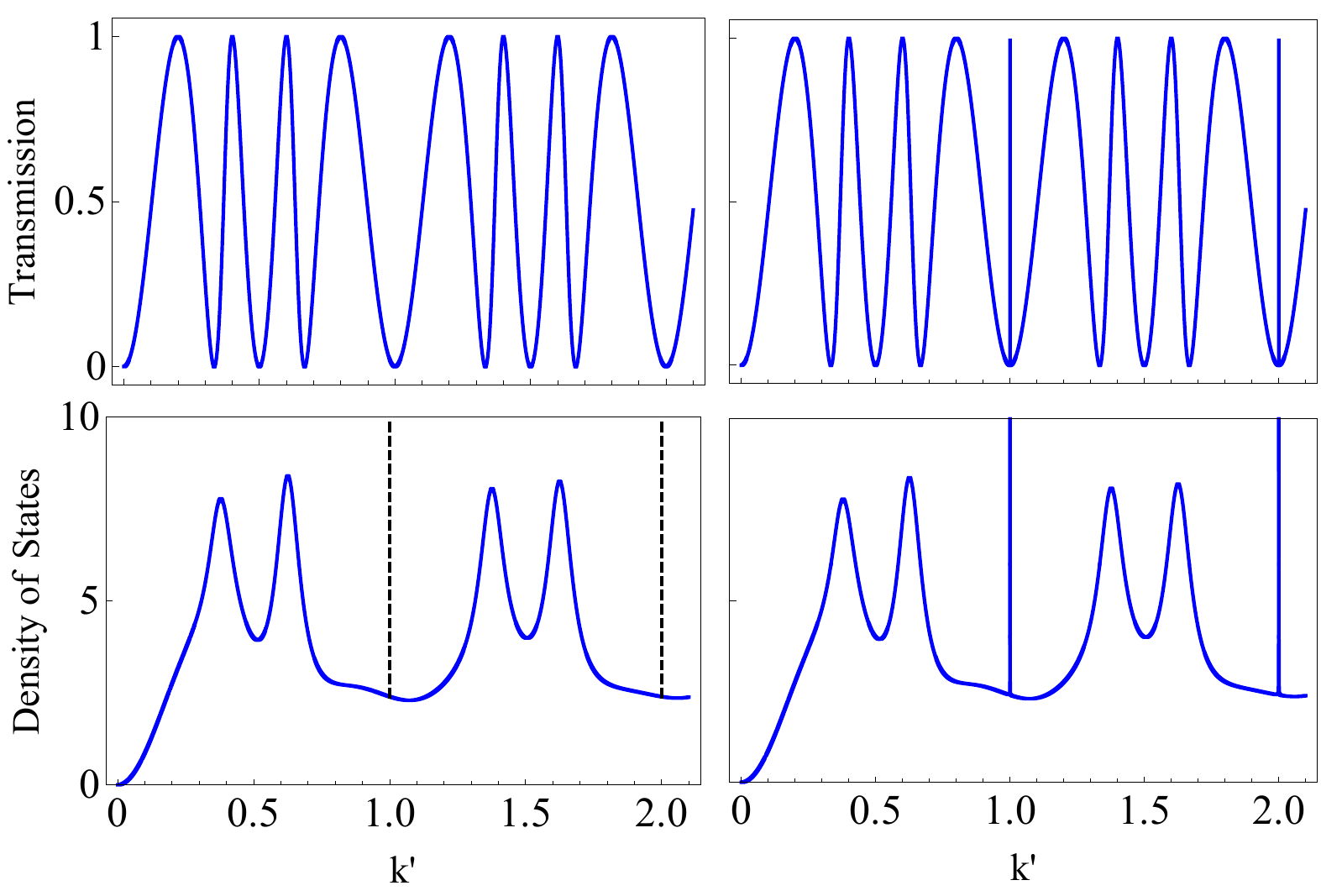}
\caption{Results for the transmission (top) and the density of states (bottom) vs. $k'$ for a single crossbar junction with $n=2$ and $m=3$. The right panels show the results with full commensurability between the upper and lower arms ($\Delta=0$), while the left panels consider a small breaking of the commensurability ($\Delta=0.001$).}
\label{Fig5}
\end{figure}

Fig. \ref{Fig5} displays transmission probability (top) and density of states (bottom) %(\ref{Fig5.1} and \ref{Fig5.2}) and the probability density (\ref{Fig5.3} and \ref{Fig5.4}) 
for  a single crossbar junction, with $n=2$ and $m=3$, for two cases: considering commensurability between the upper and lower arms by having $\Delta=0$ (right)
%(\ref{Fig5.1} and \ref{Fig5.3}) 
and assuming a rupture of this commensurability by having $\Delta=0.001$ (left).

As predicted, we can appreciate the formation of Fano resonances in every value of $k'$ that meets the condition stated in Eq. (\ref{43}) and the appearance of quasi-BICs for the rupture of commensurability by having $\Delta=0.001$ for every value of $k'$ that meets the condition stated in Eq. (\ref{44}), which leads us to recognize the formation of BICs in these same values of $k'$ for the case of absolute commensurability ($\Delta=0$).

On the other hand, computing the density of states in the system, we notice the appearance of sharp peaks for the same values of $k'$ in which they appear quasi-BICs for the case $\Delta=0.001$. These peaks correspond to Dirac's deltas with no width for the case of absolute commensurability ($\Delta=0$), so a dashed line represents them. The appearance of these Dirac's deltas confirms the formation of BICs for the values of $k'$ already mentioned.

It is possible to interpret the interesting results and the commensurability condition by considering that the vertical top and bottom arms form an infinite well. The states in the well are not affected by the continuum if there is a node of their wavefunction in the connection to the transmission waveguides. More information about this can be found in Appendix \ref{App:well}.

\begin{figure}[H]
        \centering 
        \includegraphics[width=\columnwidth]{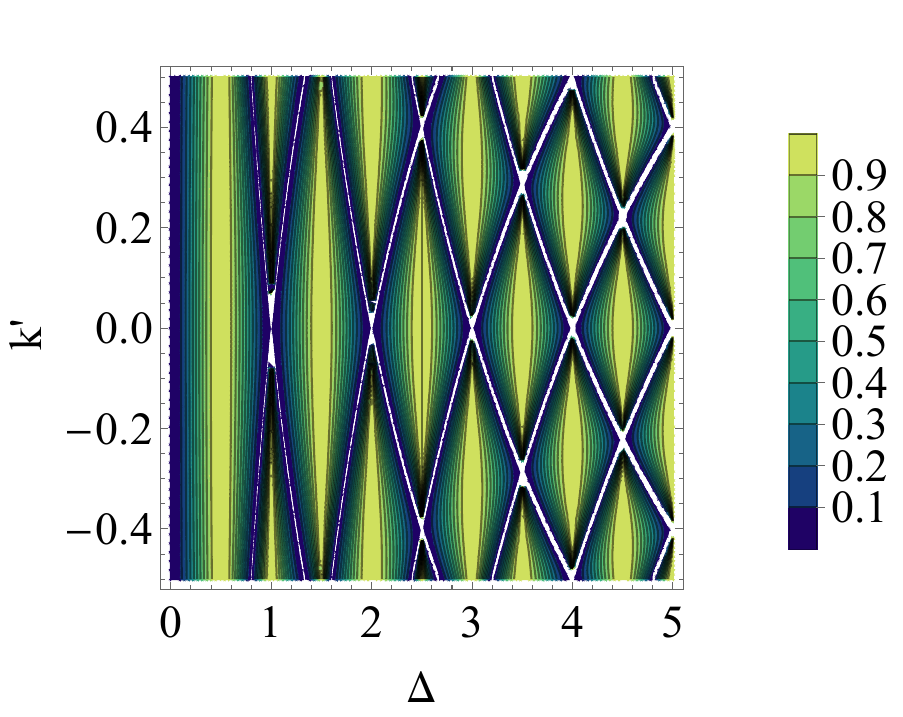}
        \caption{Contour plot for the transmission as a function of parameters $k'$ and $\Delta$ with $n=m=1$.} 
        \label{Fig:contour_single_junction}
    \end{figure}

%%%%%%%%%%%%%%%%% CREO QUE ESTO NO AÑADE MUCHA INFORMACION. NO ESTAMOS TAN INTERESADOS EN LO QUE OCURRE A Delta>> 0.001 
%%%%%%%%%%% Lo quitaría
Fig. \ref{Fig:contour_single_junction} displays a contour plot of the transmission as a function of the renormalized momenta $k'$ and the asymmetry parameter $\Delta$ for the case $n=m=1$. The plot clearly shows the evolution of the BICs into quasi-BICs and then into more standard resonances as $\Delta$ is increased.

    %%%%%%%%% QUITARIA HASTA AQUI

Now, we analyze the dependency of the quasi-BIC's width on the perturbation parameter $\Delta$. In the case of single-channel waveguides, an elastic perturbation in the arms can only change the optical path and is equivalent to the $\Delta$ parameter as defined earlier. 
For simplicity, we consider the case of $n=m=1$ without loss of generality. First, let us assume the equation for the transmission, Eq.(\ref{42}). In the limit $\Delta \ll 1$, we can write the equation of the transmission, Eq. (\ref{42}) as

\begin{equation}
    T\approx \sin^2(k'\pi)+\frac{\sin^4(\frac{\pi\Delta}{2})}{\sin^4(\frac{\pi\Delta}{2})+ \sin^2(k'\pi)}.
    \label{Eq7}
\end{equation}

We can identify, then, two different contributions to this equation. The first contribution consists of a simple quadratic Sine function and describes the interference effects and Fano resonances that appear for $\Delta=0$. The second contribution has the form of a Breit-Wigner resonance ($T(x)=\Gamma^2/(\Gamma^2+x^2)$) with width $\Gamma=\sin^2(\pi\Delta/2)$ which for small values of $\Delta$ is just $\Gamma\approx(\frac{\Delta}{2})^2\pi$.
\\

\begin{figure}[H]
        \centering 
        \includegraphics[width=0.9\columnwidth]{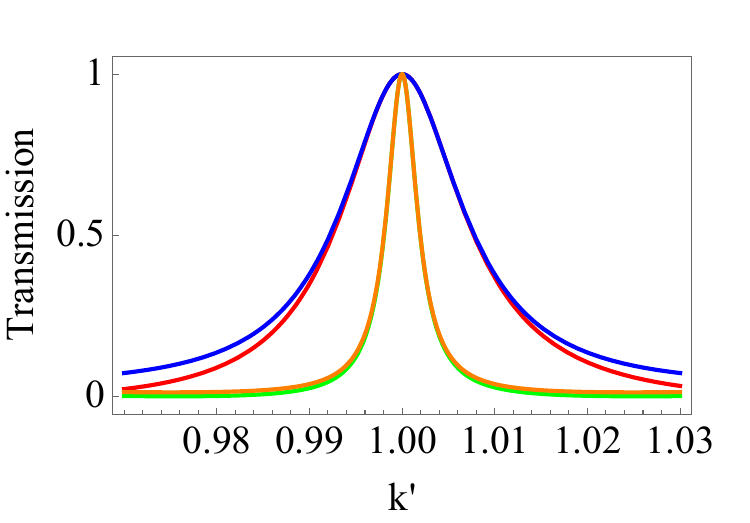}

        \caption{Transmission vs. $k'$ for a single crossbar junction with $n=m=1$. Red and green lines show the exact results from Eq. (\ref{42}) considering $\Delta=0.05$ and $\Delta=0.1$ respectively, while blue and orange lines show the approximation given by Eq. (\ref{Eq7}).  For these values of $\Delta$, both curves overlap.}
        \label{Fig111}
    \end{figure}

From the above equation, we can see that, for small perturbations $\Delta \ll 1$, the width of the quasi-BIC is proportional to the quadratic value of the perturbation parameter ($\Delta^2$). This formula should be helpful for the use of the system's BICs in metrological and sensing applications. In Fig. \ref{Fig111}, we show a comparison of the exact result with the Breit-Wigner approximation for two values of the perturbation, $\Delta=0.05$ and $\Delta=0.1$ showing a good agreement even for no so small values of $\Delta$.
\\

\begin{figure}[H]
    \includegraphics[width=1\columnwidth]{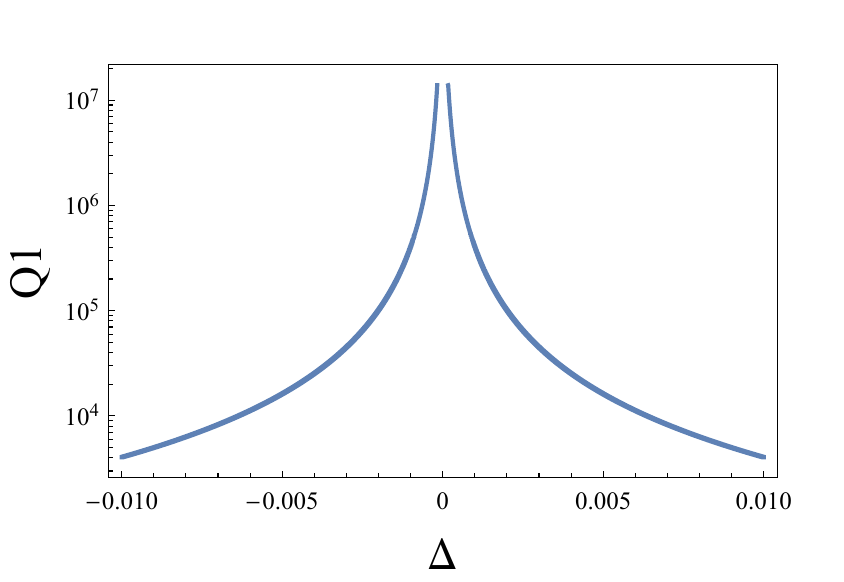}
%    \begin{justify}
    \caption{Q-factor as a function of asymmetry parameter $\Delta$ for $N=1$ considering $n=m=l=1$.}
%    \end{justify}
        \label{FigQ}
    \end{figure}

    Using Eq. (\ref{Eq7}), we can gather the data needed to calculate the Q-factor for this setup considering $n=m=l=1$. The Q-factor is represented by $Q_1=(n\pi)/\Gamma$. The Q-factor graph is shown in Fig. (\ref{FigQ}), with the asymmetry parameter ($\Delta$) as the independent variable. As we can see, this setup can achieve an ultra-high Q-factor that diverges at the resonances.
%\begin{figure}[H]
%    \begin{subfigure}
%    {0.5\textwidth}
%    \includegraphics[scale=0.5]{cb05.png}
%    \caption{$\Delta=0.05$}
%    \label{Fig11.1}
%    \end{subfigure}
%    \begin{subfigure}
%    {0.5\textwidth}
%    \includegraphics[scale=0.5]{cb1.png}
%    \caption{$\Delta=0.1$}
%    \label{Fig11.2}
%    \end{subfigure}
%    \caption{\revision{Transmission vs. k for a single crossbar junction with $n=m=1$. The exact result (solid black line) from equation \ref{42} and the approximation (dashed red line) given by equation \ref{60}. \ref{Fig11.1} Considers a perturbation parameter $\Delta=0.05$. \ref{Fig11.2} Considers a perturbation parameter $\Delta=0.1$. For these values of $\Delta$, both curves overlap}. \agregado{I would consider both cases in one figure.}}
%    \label{Fig11}
%\end{figure}

\subsection{Array of N Crossbar Junctions}

We now study a generalization of the previous system consisting of an array with N identical cross junctions, equally spaced out one from the other%, as shown in Figure \ref{Fig2}
. The separation length is given by the parameter $a$. As before, the system is open through its horizontal arms and remains closed at the ends of its vertical arms, their respective lengths $L^+$ (upper arm) y $L^-$ (lower arm).
    
%    \begin{figure*}[t]
%        \centering 
%        \includegraphics[width=1.8\columnwidth]{Ncruceschico.png}
        %\includegraphics[width=\textwidth]{Fig1.pdf}
%        \caption{System consisting of a periodic array of \emph{N} identical crossbar junctions, equally spaced out one from the other and formed by one-dimensional waveguides. The system is open through its horizontal arms and remains close at the ends of its vertical arms, being their respective lengths $L^+$ (upper arm) and $L^-$ (lower arm)}
        \label{Fig2}
%    \end{figure*}
    
We use the transfer matrix method \cite{markos2008wave} to find the transmission probability for this system as explained in Appendix \ref{App:Narray}. The final result is
\begin{equation} 
        T_N (k)=\frac{1}{1+[\frac{\abs{\alpha} sin(Nql)}{2sin(ql)}]^2},
        \label{Eq8}
    \end{equation}
with the parameter $\alpha=i\left[\cot(kL^+)+\cot(kL^-)\right]$ and 

   \begin{equation} \label{26}
        cos(ql)= cos(ka)-\frac{[cot(kL^+)+cot(kL^-)]sin(ka)}{2}.
    \end{equation}

%\subsection{Array of N Crossbar Junctions}

To analyze the transmission through the junction array and the formation of BICs in the system, we make the variable changes in Eqs. (\ref{39}) and (\ref{40}) plus the following:
\begin{equation} \label{41}
      a=l + \nu,
      %%%% PORQUE ERA /pm ANTES?
  \end{equation}
where the new parameter $l$ is also an integer. The parameter $\nu$ is a commensurability parameter that will play a similar role to $\Delta$ but for the central region.

A careful examination of Eq. (\ref{Eq8}) reveals that for some particular highly symmetric cases there is perfect transmission independently of the incident momentum. We do not explore these cases further in this work as our focus is on BICs.

\begin{figure}[H]

    \centering
    \includegraphics[width=\columnwidth]{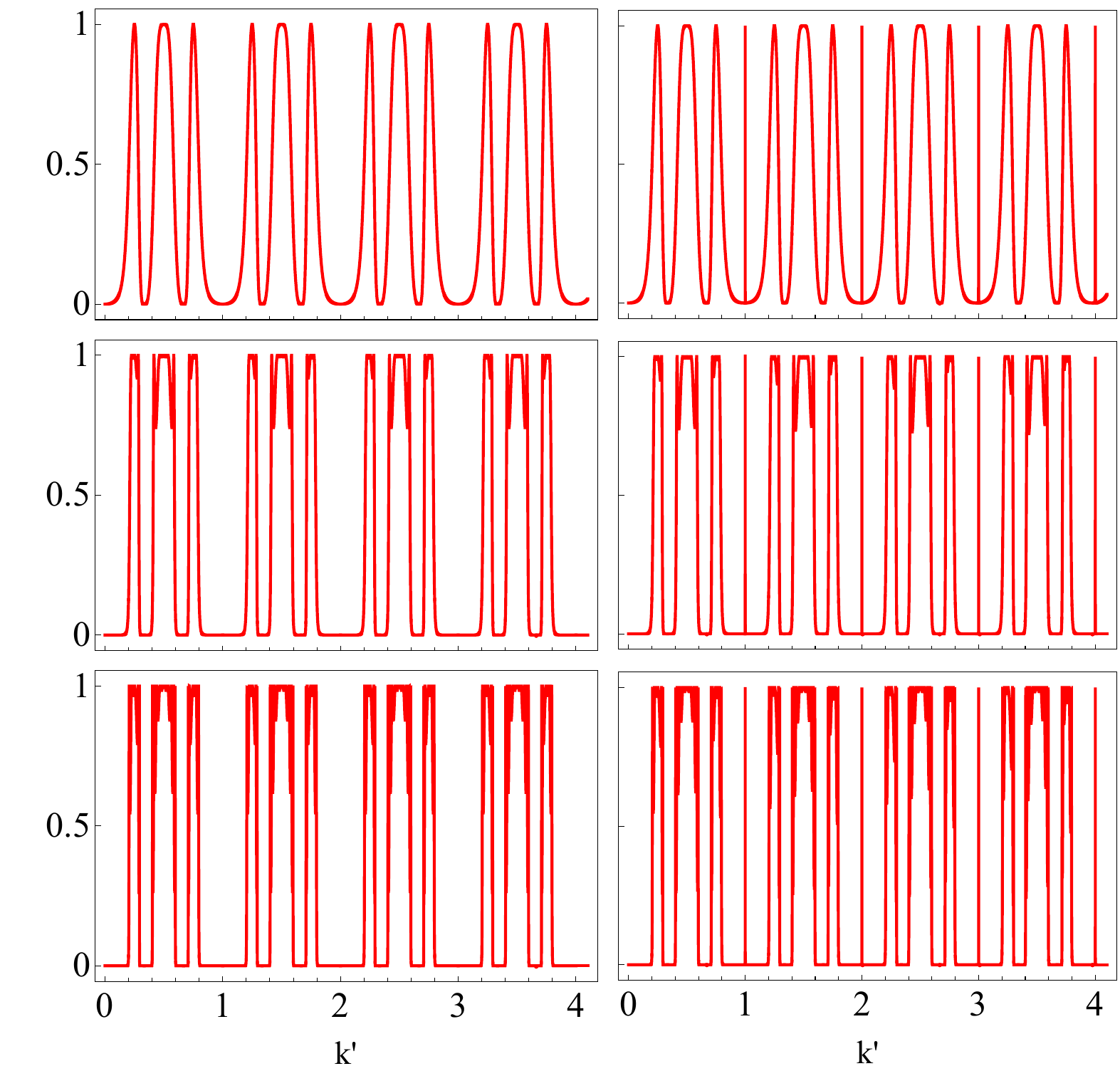}      
    \caption{Transmission vs. $k'$ for a system formed by different numbers of crossbar junctions, for $ n = 1$, $m = 3$, $l = 5$ and $\nu=0$. From top to bottom, the three presented rows show the cases with $N=2$, $N=5$, and $N=10$, respectively. Left panel shows the results for $\Delta=0$, and right panel shows the results for $\Delta=0.01$.}
    \label{Fig6}
\end{figure}

Fig. \ref{Fig6} displays the transmission profile for an N-junction system but with different values of $N$. From top to bottom, we show $N=2$, $N=5$, and $N=10$. We have chosen $ n = 1$, $m = 3$, $l = 5$ and $\nu=0$ for all cases. We compare the cases with $\Delta=0$ (left panels) with those with $\Delta=0.01$ (right panels). We observe the progressive formation of a band structure as the value of $N$ is increased. This band structure replaces the pattern shown in the transmission of resonances and anti-resonances in the single cross junction. When $\Delta \ne 0$, new narrow bands from the quasi-BICs appear in the middle of the forbidden region with almost zero transmission. 

\begin{figure}[H]
    \includegraphics[width=\columnwidth]{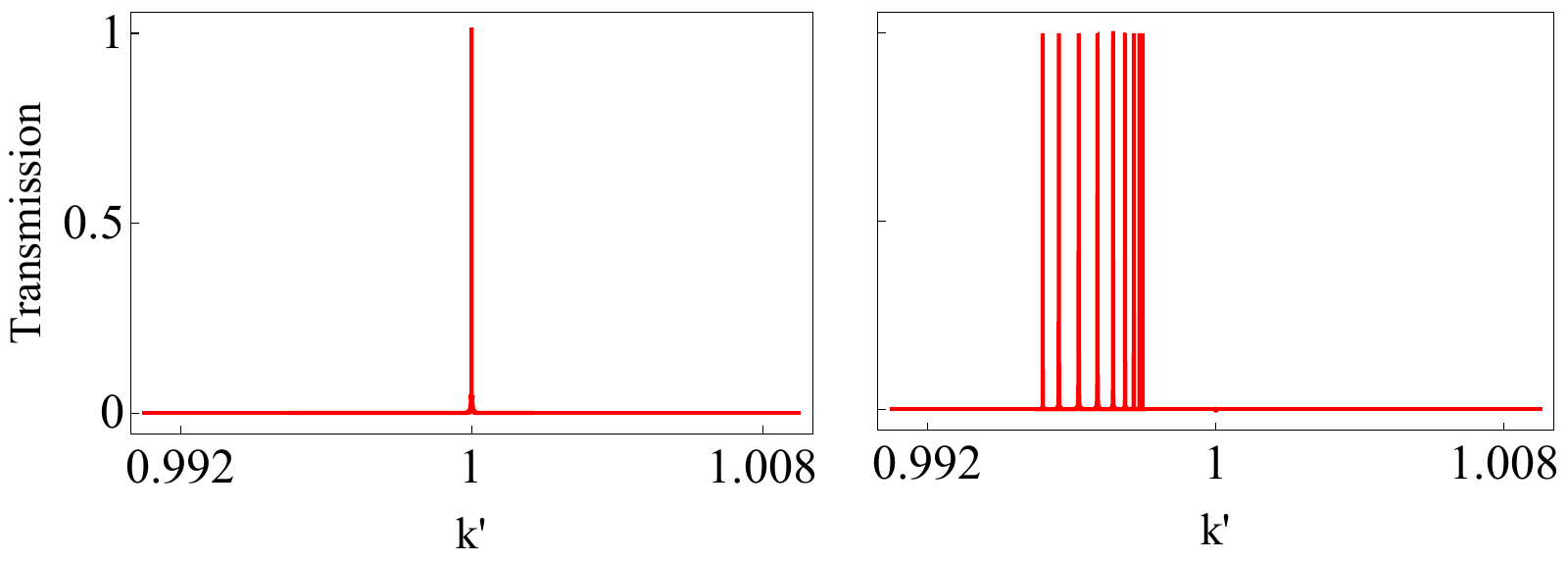}
   % \caption{$n=1$, $m=2$, $l=4$ ,$\Delta=0.001$ and $\nu=0$}
    
    \caption{Zoom of the first quasi-BIC structure formed for $N=10$ crossbar junctions with $ n = 1$, $m = 3$ and $l = 5$. Left figure shows $\Delta=0.01$ and $\nu=0$ while the right figure shows $\Delta=0$ and $\nu=0.01$.}
    %\agregado{There seems to be a lack of resolution in the figures b,c and d.}
    \label{Fig65}
    \end{figure}

Figure \ref{Fig65} shows zooms of the band structures formed for $N=10$ crossbar junctions with  $ n = 1$, $m = 3$, and $l = 5$ around $k'=1$, where the formation of the first quasi-BICs occurs. From the images, it is apparent that the quasi-bound states in the continuum (quasi-BICs) which were initially thought to be single sharp resonances, can actually be a series of peaks that are extremely close to each other, as in the case shown in the right figure of the panel. Eventually, as commensurability becomes full, all of these quasi-BICs merge into a single BIC. Increasing the number of cross junctions in the system leads to a rise in the number of maxima formed when commensurability is broken. With proper calibration and depending on the spectral resolution of the detectors, these crossbar junction arrays could be employed to enhance sensitivity for BICs applications in sensing and metrology.

%BICs will be formed when the lengths from the upper and lower arms of the system are commensurable, as before. But also BICs will appear when one of the arms is commensurable with the new separation length between consecutive intersections. In general, from Eq. (\ref{38}), we can see that BICs will be formed for every value of $k'$ that meets the following condition:

BICs occur when the lengths from the upper and lower arms of the system are commensurable, as previously mentioned. Additionally, BICs will also appear when one of the arms is commensurable with the new separation length between consecutive intersections. Generally, based on Equation (\ref{38}), BICs will be formed for every value of $k'$ that satisfies the following condition:
    \begin{equation} \label{45}
       \exists \{j_1,j_2\}\subset \{n,m,l\}\hspace{0.2cm}\land \hspace{0.2cm} \exists \{ s,p \} \in \mathbb N  \hspace{0.2cm}:\hspace{0.2cm}   k'=\frac{s}{j_1}=\frac{p}{j_2}.      
    \end{equation}  
    
Thus, if any two of these three lengths ($n$, $m$, and $l$) are commensurate, there will be a BIC at such a value of $k'$. Considering this condition and the discussion in Appendix \ref{App:well},  it is evident that the well states given rise to BICs due to its nodal structure can be the same as before or can be hybridized states between the central region and either the top or the bottom sidearm. The commensurate condition with $a$ implies that the stationary states formed by this hybridization have nodes at $x=(j-1)a$ and $x=ja$ besides canceling at $y=L^+$ or $y=L^-$. These hybridized states are the continuum equivalent of compact localized states in lattice models \cite{Maimati17}. Similarly, as these compact localized states do in infinite systems, they form BIC flat bands that became quasi-BIC very narrow bands under infinitesimal perturbations.

Figure \ref{FigDOS} displays the density of states for the system with $N=2$, $n=1$, $m=3$, and $l=5$ ( See in   Appendix \ref{App:2array} the analytic calculations).  From top to bottom, the three presented rows show the density of states for the arms of the first crossbar (entrance), middle section, and the arms of the second crossbar (exit), respectively.  The left panel shows the results for $\Delta=\nu=0$, the middle panel for $\Delta=0.001$ and $\nu=0$, and the right panel for $\Delta=0$ and $\nu=0.001$.  Once again, the appearance of Dirac's delta functions (with no width for the case of full commensurability) for the same values of $k'$ in which they appear sharp resonances in the transmission profile which we identified as quasi-BICs for the case $\Delta=0.01$ and $\nu=0$ or otherwise ($\Delta=0$ and $\nu=0.01$) confirms the formation of BICs for the values of $k'$ already mentioned.%\begin{widetext}

%{\color{red}For instance, for $N=2$, Eq. \ref{Eq8} reduces to,
%\begin{widetext} 
%\begin{equation} 
 %       T_2(k) =\frac{1}{1+[cot(kL^+)+cot(kL^-)]^2(cos(ka)-\frac{[cot(kL^+)+cot(kL^-)]sin(ka)}{2})^2}.
  %      \label{Eq12}
%   $$ \end{equation} 
%\end{widetext}
%For $n=m=l$ and $a=l+\nu$, with $\nu \ll 1$, Eq.(\ref{Eq12}) can be written as a superposition of two contributions
%\begin{equation} 
 %       T_2(k) \approx \frac{sin(kl)^2}{sin(kl)^2+16 cos(kl)^4}+\frac{\gamma^2}{(k-k_o)^2+\gamma^2}.
  %      \label{Eq13}
   % \end{equation}   
%The first term gives the Fano-like resonances, where the transmission vanishes around $kl=\pi l$ and goes to unity around $kl= \pi(l+1/2)$. The second term represents a Breit-Wigner line shape with resonances in $k_0=2\pi l/(2l+\nu)$ and width $\gamma=\frac{sin(2\pi l/(2l+\nu))^2}{8 \pi l}$. 

%Using Eqs.\ref{Eq7} and \ref{Eq13}, we can gather the data needed to calculate the Q-factor for the setups with $N=1,2$. The Q-factors are represented by $Q_1=(n\pi)/\Gamma$ and $Q_2=k_0/\gamma$, respectively. The Q-factor graph for $N=1,2$ is shown in Fig. (\ref{Q-factor}), with the asymmetry parameter ($\Delta,\delta$) as the independent variable. As we can see, both setups can achieve ultra-high Q-factors that diverge at the resonances.}
 Based on the literature \cite{Hsu16, Sadreev2021}, BICs are formed by several mechanisms, such as the symmetry-protected and Fabry-Perot mechanisms. The latter occurs when two separated resonators have a perfect reflection. Our study shows that the single cross-bar junctions extend the symmetry protection mechanism to commensurate lengths. However, in junction arrays, new BICs appear depending on the size of the central region, and these can be attributed to the Fabry-Perot mechanism. This structure is similar to the one analyzed in reference \cite{Nabol2022}.

%[PRB 106, 245403 (2022)], \cite{Bulgakov2006}.}

%\begin{figure}[!h]
 %   \includegraphics[width=1\columnwidth]{Qfactor.pdf}
   % \caption{Q-factor as a function of asymmetry parameter for $N=1$ (upper panel) and $N=2$ (lower panel) }
 %   \label{Q-factor}
  %  \end{figure}
 
\begin{widetext}
    
 \begin{figure}[H]
    \includegraphics[width=\columnwidth]{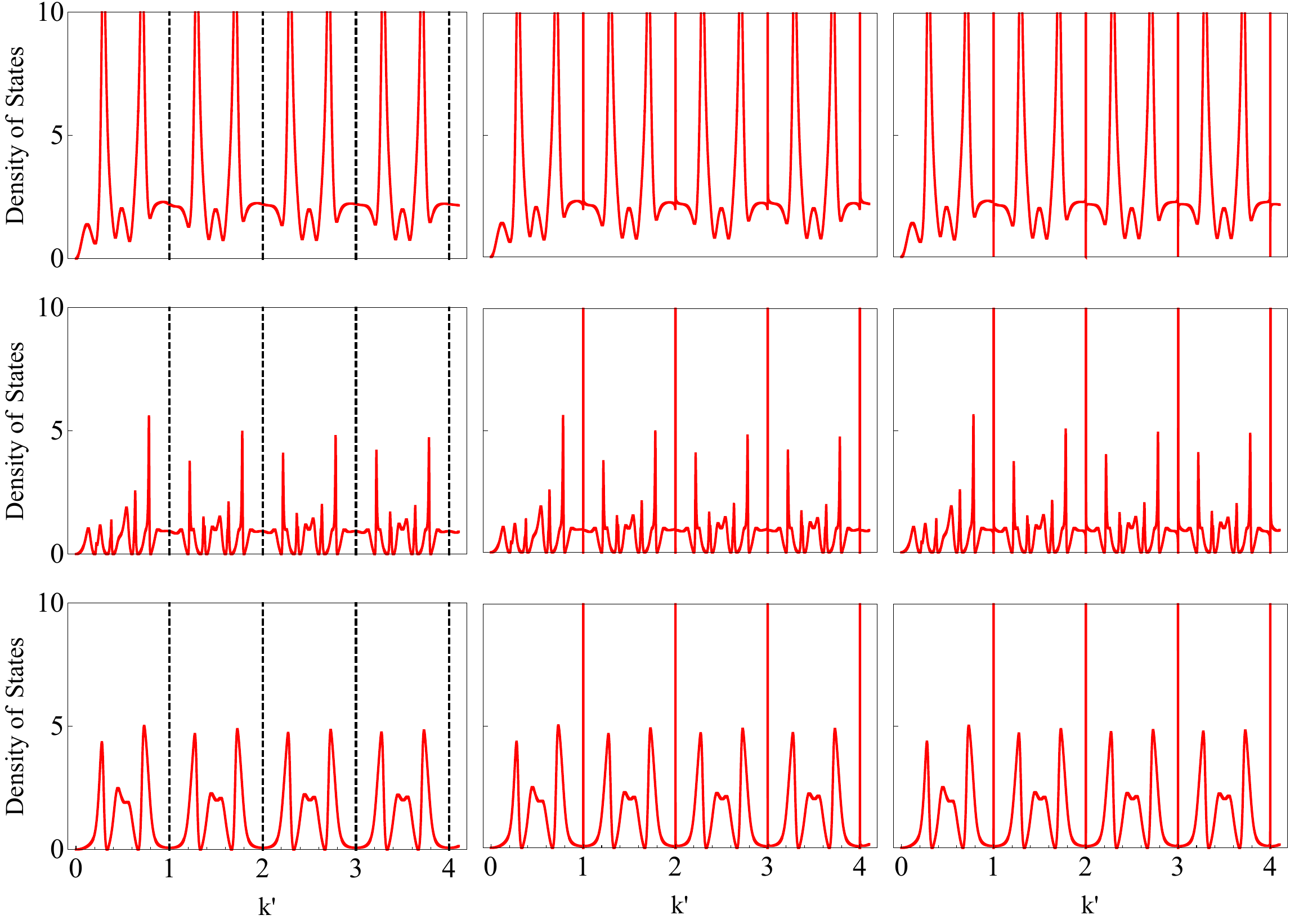}
   % \caption{$n=1$, $m=2$, $l=4$ ,$\Delta=0.001$ and $\nu=0$}

    \caption{\small{Density of states vs. $k'$ for a system formed by $N=2$ crossbar junctions, for $ n = 1$, $m = 3$ and $l = 5$. From top to bottom, the three presented rows show the density of states for the arms of the first crossbar (entrance), middle section, and the arms of the second crossbar (exit), respectively. The left panel shows the results for $\Delta=\nu=0$, the middle panel for $\Delta=0.001$ and $\nu=0$, and the right panel for $\Delta=0$ and $\nu=0.001$.}}   
    %\agregado{There seems to be a lack of resolution in the figures b,c and d.}
    \label{FigDOS}
    \end{figure}
\end{widetext}

\begin{figure}[H]
\centering
    \includegraphics[width=1\columnwidth]{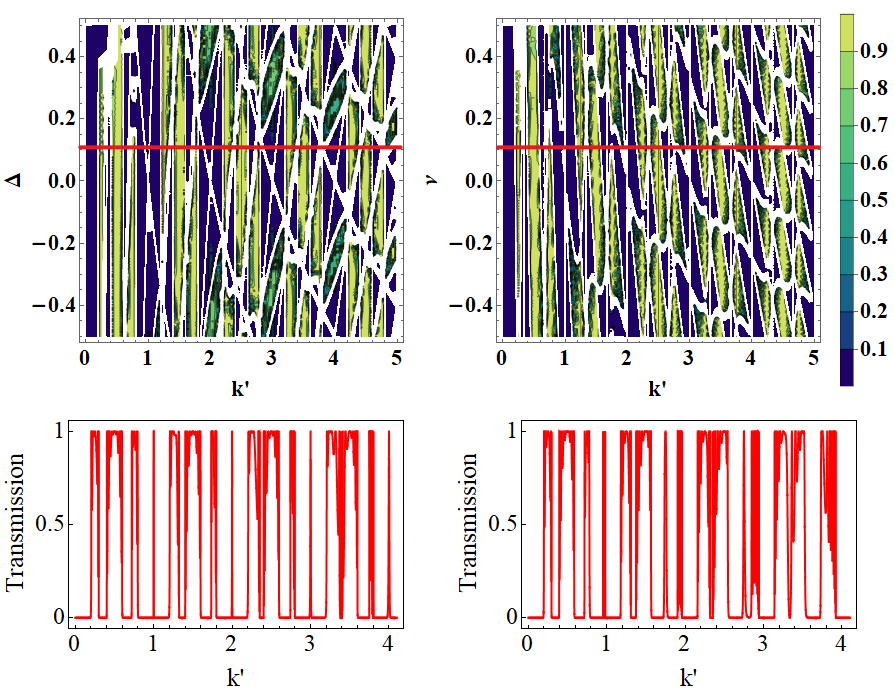}
    \caption{\small{Top panels show contour plots for the transmission with $ n = 1$, $m = 3$, $l = 5$  and $N=10$ as a function of $\Delta$ and $k'$ with $\nu=0$ (Left) and $\nu$ and $k'$ with $\Delta=0$ (Right). Red lines mark specific sections of these contour plots shown in the bottom panels with the corresponding transmission plots as a function of $k'$ with $\Delta=0.1$, $\nu=0$ (Left) and $\Delta=0$, $\nu=0.1$ (Right).}}
    \label{Fig66}
    \end{figure}

Fig. \ref{Fig66} shows a contour plot of the transmission for a system with $N=10$, $ n = 1$, $m = 3$ and $l = 5$. In the left panels, we show the case with $\nu=0$ as a function with $\Delta$ and $k'$, while in the right panels, we show the case with $\Delta=0$ as a function of $\nu$ and $k'$. The parameter $\Delta$ is symmetric between positive and negative values as the change of sign only switches the role of the upper and lower arms of the junction. However, the parameter $\nu$ shows a clear asymmetry between positive and negative values as larger or smaller central regions change the commensurability conditions with the arms differently. 
    
\begin{figure}[H]
    \includegraphics[width=\columnwidth]{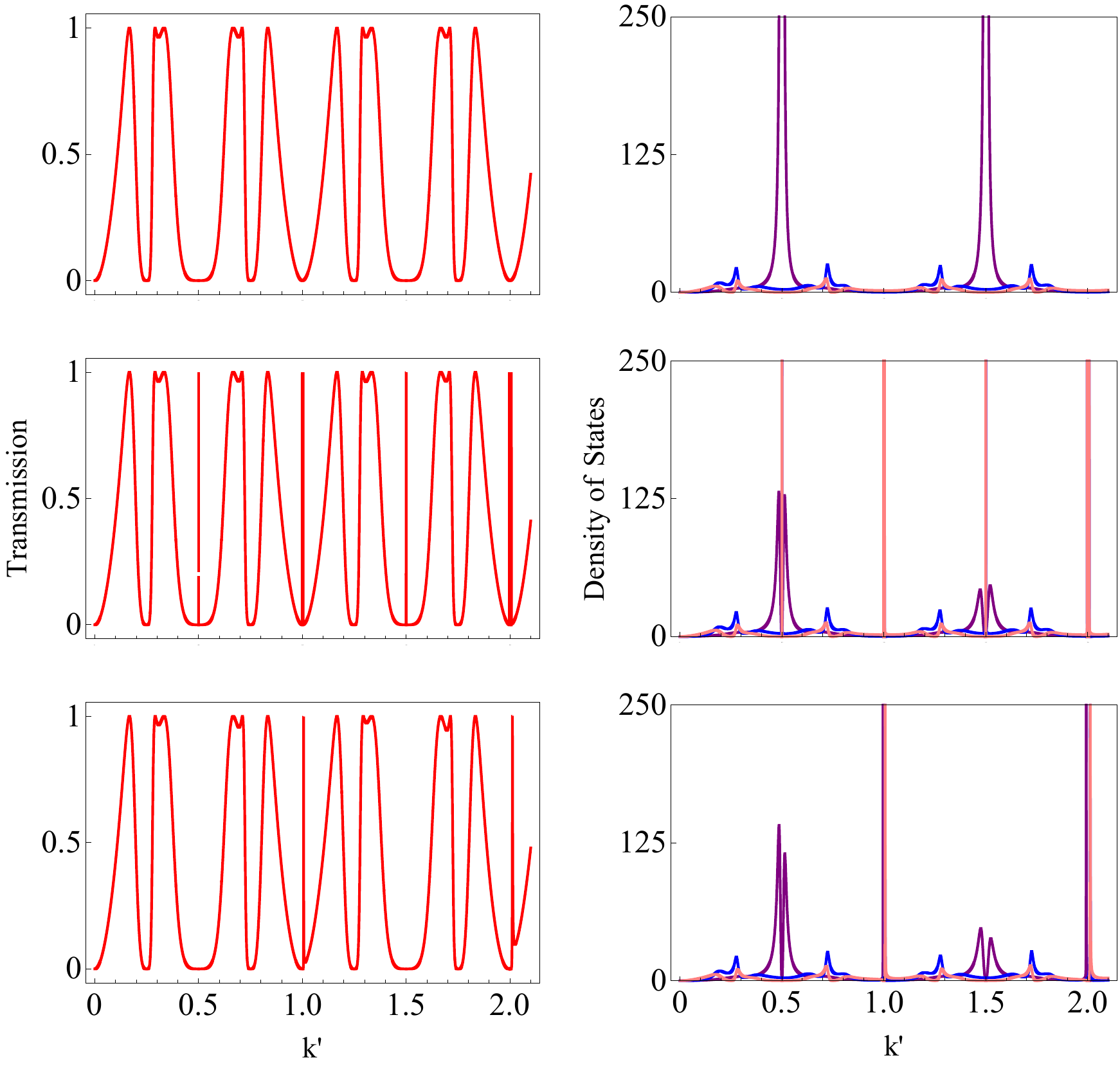}
    %\caption{$n=1$, $m=2$, $l=4$ ,$\Delta=0$ and $\nu=0$}
    \caption{\small{System formed by $N=2$ crossbar junctions with measures $n=4$, $m=2$, and $l=1$. The left panels correspond to the transmission, while the right panels show the density of states for different regions separated by colors: blue lines for the arms in the first crossbar, purple for the middle section, and pink for the arms in the second crossbar. From top to bottom, the three presented rows correspond to $\Delta=\nu=0$, $\Delta=0.01$ with $\nu=0$, and finally $\nu=0.01$ with $\Delta=0$.} }   \label{Figgg}
    \end{figure}

Figure \ref{Figgg} displays the transmission and density of states of a system consisting of two crossbar junctions with dimensions of $n=4$, $m=2$, and $l=1$. The colors of the curves represent the local density of states in three different regions: the arms of the first crossbar, the middle region, and the arms of the second crossbar. According to expression \ref{45}, we expect BICs to form for values of $k'=\frac{s}{2}$ with $s$ an integer. Considering the formation of quasi-BICs shown in the central and bottom panels (where the system's commensurability has been broken with $\Delta=0.01$ and $\nu=0.01$, respectively), we confirmed the formation of BICs for the expected values. Depending on the breaking that we have in the system's commensurability, different quasi-BICs form. In the central panels the rupture occurs between ($n$ and $m$), between ($n$ and $l$) and between ($m$ and $l$), so quasi-BICs are formed in every value of $k'=\frac{s}{2}$. These quasi-BICs show in the density of states profile as a superposition of very narrow peaks %Dirac's deltas 
for the different regions of the system. In this case, quasi-BICs are formed in the arms of both crossbar junctions but not in the middle section (this is a superposition of blue and pink almost Dirac's delta lines 
in the profile). On the other hand, in the bottom panels, the rupture occurs only between ($n$ and $l$) and ($m$ and $l$). However, commensurability between ($n$ and $m$) remains unbroken. That's why quasi-BICs form for only integer values of $k'$ as $k'=\frac{s}{1}$ with $s \in \mathbb{N}$. In this case, quasi-BICs are formed in both crossbar junctions and in the middle section (this is a superposition of blue, pink and purple narrow peaks
%Dirac's delta lines 
in the profile).

%Figure \ref{Fig8} shows a zoom of the quasi-BICs formed by breaking commensurability for the case $n=1$, $m=3$, and $l=5$. As seen from the images, those quasi-BICs that seemed like single sharp resonances are a set of peaks very close to each other. Eventually, all these quasi-BICs collapse into a BIC for full commensurability. The amount of maxima formed by breaking commensurability increases as we increase the number \emph{N} of cross junctions in the system. After proper calibration and depending on the spectral resolution of the detectors, these arrays of crossbar junctions could be used to enhance sensitivity for BICs applications in sensing and metrology.

%\begin{figure}[H]
    %\includegraphics[width=\columnwidth]{panel 5 con eje.pdf}
   % \caption{$n=1$, $m=2$, $l=4$ ,$\Delta=0.001$ and $\nu=0$}
    
%    \caption{Zoom of the first quasi-BICs formed for $N=5$ crossbar junctions with $n=1$, $m=3$ and $l=5$. Left figure shows $\Delta=0.01$ and $\nu=0$ while the right figure shows $\Delta=0$ and $\nu=0.01$. We notice that the figures that appeared to be single quasi-BICs for the rupture of commensurability are a group of sharp resonances very close to each other.}
    %\agregado{There seems to be a lack of resolution in the figures b,c and d.}
%    \label{Fig8}
 %   \end{figure}
    
%In Appendix \ref{App:Double} we also study explicitly the interesting example of $N=2$.
    
\section{T-shape junctions}

A T-shape junction is a particular case of our crossbar junction model with one of the lengths going to zero. For example, let us put $L^-=0$, and the transmission formula is then simplified to:
 \begin{equation} \label{eq:Ttshape}
    T=\abs{t}^2= \frac{4}{4+[cot(kL^+)]^2}.
    \end{equation}
All allowed momenta in the infinite well of length $L^+$ become BICs and, with the condition for BIC formation, becomes $k_j=j \frac{\pi}{L^+}$. A change in the length of the sidearm does not induce a quasi-BIC but moves the position of the BIC in momenta. Recently, BICs in this kind of structure were explored experimentally in radio-frequency circuits \cite{Khattou2023}.

In the case of the T-shape junctions array, we recover a commensurability condition between the length of the sidearm $L^+$ and the central region between two consecutive junctions $a$. The results can also be obtained as the limit for the crossbar junction arrays with $L^-=0$. The commensurability condition for the existence of BICs becomes then:

\begin{equation} \label{Eq:BICTshape}
       \exists(s,p)\in \mathbb N \hspace{0.5cm}:  \hspace{0.5cm} 
       k'=\frac{s}{n}=\frac{p}{l}. \hspace{0.5cm}
    \end{equation}
Then the states of the system giving rise to BICs are hybridized states between the sidearm and the central region. 

\section{Summary and conclusions}
%\subsection{\revision{Conclusions}}
In summary, we studied the formation of BICs in a system consisting of crossbar and T-shape junctions in one-dimensional waveguides. To solve the problem, we use the transfer matrix method. We calculate the transmission spectrum and the density of states of the structure. In the first place, we investigate a single crossbar junction and find the formation of BICs, even in high asymmetric structures, as long as the upper and lower arms of the crossbar junctions have commensurable lengths. In the second place, we show another mechanism for forming BICs, due to the hybridization of the states in the region inter crossbar or T-shape junction with the states of the arms of the junctions. Besides, we identify a third mechanism for forming BICs by indirectly coupling two junction levels through the common unidimensional channel. An important aspect of our work is that the simplicity of the studied models allows for complete analytical treatment and a good understanding of the mechanisms behind forming such elusive states as the BICs. 

In conclusion, we characterized the formation of BICs in 
%highly 
asymmetric systems, with the condition of commensurability between the different sections of the structure.
Furthermore, another mechanism that contributes to forming BICs, which is connected to the Fabry-Perot mechanism, has been identified. This mechanism uses two consecutive crossbar structures as perfect mirrors, generating stationary waves within the effective cavity that decouple from the continuum.
Our results could be applied in developing sensors as we have studied how the width and the Q-factor of the quasi-BICs depend on a perturbation. Our results go beyond the symmetry-protected BIC paradigm and may inspire a new search for BICs and their applications in different wave systems. We expect the BICs reported in this work to be observed in many different classical and quantum wave systems. An interesting setup to observe these states would be as standing waves in acoustic waveguides with resonant cavities, based on the work carried out by Huan {\em et al.} \cite{huang2021sound}. A simple extension of the experimental setup by Khattou {\em et al.} with stubbed structures in coaxial cables working in the radio-frequency domain \cite{Khattou2023} should allow us to test our results.
Moreover, our work raises new questions to answer in further investigations and the possibility of interesting extensions. What would be the role of disorder in the case of the junction arrays? What happens if we increase the dimensionality of the arrays? Can the results be extended to multichannel systems? The condition for BICs will probably depend on the mode, but this dependency may have interesting applications for mode selectivity. Engineering the lengths and couplings in our models, it should be possible to induce non-trivial topology in the arrays as has already been done in different photonic systems \cite{Siroki2017,Khattou2023}. It would be exciting to check whether the analytical results can be obtained for non-trivial topological systems. It would also be interesting to study the anomalous highly symmetric cases mentioned earlier for the array of N identical cross junctions with perfect transmission. We plan to explore all these aspects in future works. 

\begin{acknowledgments}
We acknowledge financial support by Projects No. PGC2018-094180-B-I00, PID2019-106820RBC21 and PID2022-136285NB
funded by MCIN/AEI/10.13039/501100011033 and FEDER ”A way of making Europe”, ANID-Subdirección de Capital Humano/Magíster Nacional/2023-22231536, PIIC No. 050/2023 and FONDECYT Grant 1201876.
\end{acknowledgments}

\appendix

\section{Analytic solution of the single crossbar junction}
\label{AppA}

In this appendix, we present the analytic solution for the scattering problem of the single crossbar junction as presented in Fig. \ref{Fig1}.

We separate the system into four sections: the left arm will be section 1, the right arm section 2, the upper part section 3, and the lower part section 4. Each of these sections has an associated wave function. We consider an incident wave plane from the left, reflecting with amplitude r and transmitting with amplitude $t$.

    \begin{equation} \label{1}
    \Psi_1(x)=e^{ikx}+re^{-ikx}
    \end{equation}

    \begin{equation}\label{2}
    \Psi_2(x)=te^{ikx}
    \end{equation}

    \begin{equation}\label{3}
    \Psi_3(y)=Ce^{iky}+De^{-iky}
    \end{equation}

    \begin{equation}\label{4}
    \Psi_4(y)=Fe^{iky}+Ge^{-iky}
    \end{equation}

We seek to determine the amplitudes $r,t,C,D,F,G$, for which we must apply the corresponding boundary conditions \cite{griffiths2018introduction}. The wave function must be continuous at every point, so considering the junction point $x=y=0$, the first condition to be fulfilled is:

\begin{equation}\label{5}
\begin{split}
    \Psi_1(0)=\Psi_2(0)=\Psi_3(0)=\Psi_4(0)\\
    \Longrightarrow t=1+r=F+G=C+D
    \end{split}
\end{equation}
Secondly, we assume that wave functions at the ends of both vertical arms (sections 3 and 4) go to zero, which mathematically implies:
    \begin{equation} \label{6}
    \Psi_3(L^+)=0\Longrightarrow D=-Ce^{2ikL^+}
    \end{equation}

    \begin{equation} \label{7}
    \Psi_4(-L^-)=0\Longrightarrow F=-Ge^{2ikL^-}
    \end{equation}
    
The last condition states that the difference between the derivatives of the wave functions in the horizontal direction and those of the vertical direction must take the same value at the intersection point:
    \begin{equation} \label{8}
    \left.\pdv{\Psi_2}{x}\right\rvert_{0^{+}}-\left.\pdv{\Psi_1}{x}\right\rvert_{0^{-}}=\left.\pdv{\Psi_3}{y}\right\rvert_{0^{+}}-\left.\pdv{\Psi_4}{y}\right\rvert_{0^{-}}.
    \end{equation}
These conditions impose the following relation among the wave function coefficients in the different regions:
\begin{equation} \label{eq:coeff_condition}
   t+r-1=C+G-D-F 
\end{equation}
    
We must solve for coefficients \emph{r} (reflection amplitude), \emph{t} (transmission amplitude), C, D, F, and G by using equations \ref{5} to \ref{eq:coeff_condition}. The explicit solution can be written as follows: 
%for being able to calculate the total probability of transmission as: %Solving the equation system we now have: 
    \begin{equation} \label{9}
   t=\frac{2}{2-i[cot(kL^+)+cot(kL^-)]}
    \end{equation}
    \begin{equation} \label{10}
    r=\frac{2}{2-i[cot(kL^+)+cot(kL^-)]}-1
    \end{equation}
    \begin{equation} \label{11}
    F=\frac{2e^{ikL^-}}{4isin(kL^-)+2sin(kL^-)[cot(kL^+)+cot(kL^-)]}
    \end{equation}
    \begin{equation} \label{12}
    G=\frac{-2e^{-ikL^-}}{4isin(kL^-)+2sin(kL^-)[cot(kL^+)+cot(kL^-)]}
    \end{equation}
    \begin{equation} \label{13}
    C=\frac{-2e^{-ikL^+}}{4isin(kL^+)+2sin(kL^+)[cot(kL^+)+cot(kL^-)]}
    \end{equation}
    \begin{equation} \label{14}
    D=\frac{2e^{ikL^+}}{4isin(kL^+)+2sin(kL^+)[cot(kL^+)+cot(kL^-)]}
    \end{equation}

From the squared transmission amplitude $|t|^2$ we can compute the total transmission as in Eq. \ref{15} in the main text.
   %\begin{equation} \label{15}
    %T=\abs{t}^2= \frac{4}{4+[cot(kL^+)+cot(kL^-)]^2}
    %\end{equation}
    
We can also find the local density of states as a function of $k$ for the upper and lower sections of the crossbar junction, which are respectively given by:
    \begin{equation} \label{16}
    P_{3}=\int_{y=0}^{y=L^+}\abs{\Psi_3}^2dy=\frac{-2cot(kL^+)+2kL^+csc(kL^+)^2}{k[4+[cot(kL^+)+cot(kL^-)]^2]}
    \end{equation}

    \begin{equation} \label{17}
    P_{4}=\int_{y=-L^-}^{y=0}\abs{\Psi_4}^2dy=\frac{-2cot(kL^-)+2kL^-csc(kL^-)^2}{k[4+[cot(kL^+)+cot(kL^-)]^2]}
    \end{equation}
    
%Now, the total probability density of the vertical region in the crossbar is given by the sum of the two results obtained in equations \ref{16} and \ref{17}:
%    \begin{equation} \label{18}
%    P_{T}=P_{3}+P_{4}=-\frac{2[cot(kL^-)+cot(kL^+)-k(L^-csc(kL^-)^2+L^+csc(kL^+)^2)]}{k[4+[cot(kL^+)+cot(kL^-)]^2]}
%    \end{equation} 

\section{Quantum well solution}
\label{App:well}

We can apply the solution of a well with infinite walls to the vertical sidearms. Taking the textbook solution for the infinite potential well \cite{griffiths2018introduction}, the allowed momenta of the stationary waves solving the problem are then:
\begin{equation}\label{eq:kj}
k_j=j\frac{\pi}{L^++L^-},
\end{equation}
with $j \in \mathbb{N}$.
The corresponding wave functions
\begin{equation}
    \Psi_j=A\sin{\left[ k_j \left( y-y_c+\frac{L^++L^-}{2}\right)\right]},
\end{equation}
where $A$ is the normalization constant and $y_c=(L^+-L^-)/2$ is the middle point of the well. It is easy to get the condition for the wave function to have a node at the connection point with the transmission lines $\Psi_j(0)=0$ when $L^+=n$ and $L^-=m$ with $\Delta=0$ as in the definitions of Eqs. (\ref{39}) and (\ref{40}).
\begin{equation}
    k=s\pi,
\end{equation}
with $s \in \mathbb{N}$, which, when taken into account the allowed momenta Eq. (\ref{eq:kj}), is equivalent to the condition written in Eq. (\ref{44}) in the main text.

\section{Analytic solution of the N-identical array of crossbar junctions}
\label{App:Narray}

In this appendix, we present in detail the analytic solution for the scattering problem of the N-array of crossbar junctions as presented in Fig. \ref{Fig2} and solved through the transfer matrix method. The transfer matrix is the matrix that connects the waves to the right of the system with the waves to the left of the system as opposed to the scattering matrix that connects outgoing waves to incoming waves \cite{markos2008wave}.

\begin{figure}[H]
       \centering 
       \includegraphics[width=0.9\columnwidth]{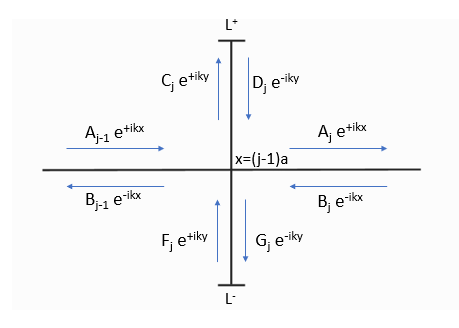}
%      %\includegraphics[width=\textwidth]{Fig1.pdf}

      \caption{jth cross junction from the array formed by \emph{N} identical cross junctions, equally spaced out one from the other. The studied junction is in position $x=(j-1)a$.}
      \label{Fig3}
\end{figure}

First, let's consider the jth cross junction from the array, placed in the position $x=(j-1)a$ 
as shown in Figure \ref{Fig3}.

The same as we did for the single junction case, we separate the system into four sections: the left arm is described by wave function $\Psi_{j-1}(x)$, the right arm by $\Psi_{j}(x)$, the upper arm by $\Phi_j^u(y)$ and the lower arm by $\Phi_j^d(y)$. In the end, considering plane wave solutions, we have:

    \begin{equation}\label{19}
    \Psi_{j-1}(x)=A_{j-1}e^{ikx}+B_{j-1}e^{-ikx}
    \end{equation}

    \begin{equation}\label{20}
    \Psi_j(x)=A_{j}e^{ikx}+B_{j}e^{-ikx}
    \end{equation}

    \begin{equation}\label{21}
    \Phi_j^u(y)=C_{j}e^{iky}+D_{j}e^{-iky}
    \end{equation}

    \begin{equation}\label{22}
    \Phi_j^d(y)=F_{j}e^{iky}+G_{j}e^{-iky}
    \end{equation}
    
We now apply the corresponding boundary conditions, which are analogous to the ones described by equations \ref{5} to \ref{8}:
    \begin{eqnarray}\label{23}
    &A_{j-1}e^{ik(j-1)a}+B_{j-1}e^{-ik(j-1)a} \nonumber \\&= A_{j}e^{ik(j-1)a}+B_{j}e^{-ik(j-1)a} \nonumber \\ &= F_{j}+G_{j}= C_{j}+D_{j}
    \end{eqnarray}
    \begin{equation} \label{24}
    \Phi_j^u(L^+)=0\Longrightarrow D_{j}=-C_{j}e^{2ikL^+}
    \end{equation}
    \begin{equation} \label{25}
    \Phi_j^d(-L^-)=0\Longrightarrow F_{j}=-G_{j}e^{2ikL^-}
    \end{equation}
    \begin{eqnarray} \label{26}
   &A_{j}e^{ik(j-1)a}-B_{j}e^{-ik(j-1)a}-A_{j-1}e^{ik(j-1)a}+B_{j-1}e^{-ik(j-1)a} \nonumber \\
   &=C_{j}-D_{j}-F_{j}+G_{j}      
    \end{eqnarray}
Let us consider, for simplicity, the following variable changes in which we have absorbed the exponential into the coefficients:
    \begin{equation} \label{27}
     A_{j}e^{ikja}=A_{j}'\hspace{1cm} \forall j
    \end{equation}
    \begin{equation} \label{28}
     B_{j}e^{-ikja}=B_{j}'\hspace{1cm} \forall j
    \end{equation}
%Considering these variable changes, 
We can now write the equation system described by equations \ref{23} to \ref{26} in its matrix form as:
    \begin{equation} \label{29}
        \begin{pmatrix} A_{j}' \\ B_{j}' \end{pmatrix}=  \begin{pmatrix} (1+\frac{\alpha}{2})e^{ika} & \frac{\alpha}{2}e^{ika} \\ -\frac{\alpha}{2}e^{-ika} & (1-\frac{\alpha}{2})e^{-ika} \end{pmatrix} \begin{pmatrix} A_{j-1}' \\ B_{j-1}'  \end{pmatrix}
    \end{equation}
For simplicity, we introduce a new parameter $\alpha$, which corresponds to:
    \begin{equation} \label{30}
        \alpha=\frac{1+e^{2ikL^+}}{1-e^{2ikL^+}}+\frac{1+e^{2ikL^-}}{1-e^{2ikL^-}}=i[cot(kL^+)+cot(kL^-)]
    \end{equation}
With this, we found the transfer matrix that represents a single cross junction of the whole system, given by:
    \begin{equation} \label{31}
        \textbf{M}_1=\begin{pmatrix} (1+\frac{\alpha}{2})e^{ika} & \frac{\alpha}{2}e^{ika} \\ -\frac{\alpha}{2}e^{-ika} & (1-\frac{\alpha}{2})e^{-ika} \end{pmatrix}
    \end{equation}
The final transfer matrix for the whole array is given by the Nth power of the $\textbf{M}_1$ matrix, that is:
    \begin{equation} \label{32}
        \textbf{M}=\textbf{M}_1^N=\begin{pmatrix} (1+\frac{\alpha}{2})e^{ika} & \frac{\alpha}{2}e^{ika} \\ -\frac{\alpha}{2}e^{-ika} & (1-\frac{\alpha}{2})e^{-ika} \end{pmatrix}^N
    \end{equation}
From the latter expression, we notice $\det \textbf{M}_1 = 1$, and we can use Chebishev's identity for computing the Nth power of the matrix.

%the eigenvalues are given by:

%    \begin{equation} \label{33}
%        \lambda=cos(ka)+\frac{i\alpha sin(ka)}{2}\pm \sqrt{[cos(ka)-\frac{\abs{\alpha}sin(ka)}{2}]^2-1}
%    \end{equation}
%For being able to apply Chebyshev's identity we need the eigenvalues to take the form:
%    \begin{equation} \label{34}
%        \lambda=e^{\pm iqa}
%    \end{equation}
%Which leads us to express cos$(qa)$ as:
%    \begin{equation} \label{35}
%        cos(qa)=cos(ka)+\frac{i\alpha sin(ka)}{2}=cos(ka)+\frac{[cot(kL^+)+cot(kL^-)]sin(ka)}{2}
%    \end{equation}  
%We can now find an expression for $qa$ and apply Chebishev's identity.\\
Chebishev's identity states that for a matrix of the form:
    \begin{equation} \label{35.1}
        \textbf{M}= \begin{pmatrix}
            a & b\\
            c & d
        \end{pmatrix}
    \end{equation}
Which eigenvalues have the form: 
    \begin{equation} \label{35.2}
        \lambda_1=e^{iq\emph{l}}
    \end{equation}
    \begin{equation} \label{35.3}
        \lambda_2=e^{-iq\emph{l}}
    \end{equation}
It's \emph{N}th power is given by:
    \begin{equation} \label{35.4}
        \textbf{M}^N= \begin{pmatrix}
            a & b\\ c & d
        \end{pmatrix}^N = \begin{pmatrix}
            a U_{N-1}-U_{N-2} & b U_{N-1}\\ c U_{N-1} & d U_{N-1}-U_{N-2}
        \end{pmatrix}
    \end{equation}
Where $ U_{N} $ is defined in function of \emph{q} as:
    \begin{equation} \label{35.5}
        U_N=\frac{\sin{(N+1)ql}}{\sin{ql}}
    \end{equation}
We can now obtain the final expression for the transfer matrix of the complete array by applying Chebishev's identity:
    \begin{equation} \label{36}
        \textbf{M}=\begin{pmatrix}
            (1+\frac{\alpha}{2})e^{ika} U_{N-1}-U_{N-2} & \frac{\alpha}{2}e^{ika} U_{N-1}\\ -\frac{\alpha}{2}e^{-ika} U_{N-1} & (1-\frac{\alpha}{2})e^{-ika} U_{N-1}-U_{N-2}
        \end{pmatrix}
    \end{equation}
Last, we can derive the transmission probability for this system from equation \ref{36}:
%    \begin{equation} \label{38}
%    \begin{split}
%        T & =\frac{1}{1+[\frac{\abs{\alpha} sin(Nqa)}{2sin(qa)}]^2}\\
%        & =\frac{1}{\frac{(\cot (k L^+)+\cot (k L^-))^2 \sin ^2\left(\text{N} \cos ^{-1}\left(\frac{1}{2} \sin (a k) (\cot (k L^+)+\cot (k L^-))+\cos (a k)\right)\right)}{4 \left(1-\left(\frac{1}{2} \sin (a k) (\cot (k L^+)+\cot (k L^-))+\cos (a k)\right)^2\right)}+1}
%    \end{split}
%    \end{equation}

    \begin{equation} \label{38}
        T =\frac{1}{1+[\frac{\abs{\alpha} sin(Nql)}{2sin(ql)}]^2}
    \end{equation}

\section{Density of states for an array of 2 identical crossbar junctions}
\label{App:2array}   

\begin{figure}[H]
        \centering 
        \includegraphics[scale=0.6]{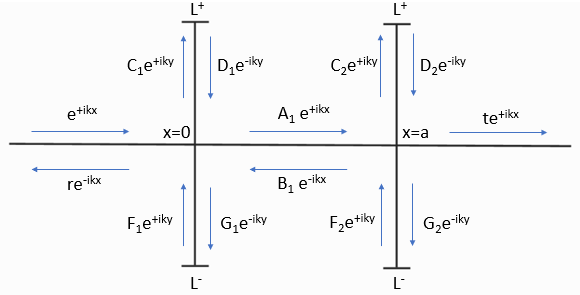}

        \caption{System consisting of two identical crossbar junctions, separated by a distance \emph{a} one from the other.}
        \label{Fig444}
    \end{figure}

When considering a system formed by $N=2$ identical crossbar junctions it's reduced to the one shown in Figure \ref{Fig444}
From Eq. (\ref{Eq8}) we express the transmission probability for $N=2$ as:
    \begin{equation} \label{39.1}
    \begin{split}
        T=\frac{1}{1+\abs{\alpha}^2[cos(ka)-\frac{[cot(kL^+)+cot(kL^-)sin(ka)}{2}]^2}
    \end{split}
    \end{equation}
We seek to determine the probability density for three regions of interest: the intermediate region between both junctions, the vertical region of the left crossbar and the vertical region of the right crossbar. For this purpose, we calculate the coefficients applying the corresponding boundary conditions shown in equations \ref{23} to \ref{26}. The probability density for the region between junctions is calculated as: 
    %\begin{equation} \label{40}
    %A=t(\frac{2-\alpha}{2})
    %\end{equation}
    %\begin{equation} \label{41}
    %B=\frac{\alpha}{2}te^{2ika}
    %\end{equation}
    %\begin{equation} \label{42}
    %C_1=\frac{t(2-\alpha+\alpha e^{2ika})}{2(1-e^{2ikL_+})}
    %\end{equation}
    %\begin{equation} \label{43}
    %D_1=\frac{-te^{2ikL_+}(2-\alpha+\alpha e^{2ika})}{2(1-e^{2ikL_+})}
    %\end{equation}
    %\begin{equation} \label{44}
    %G_1=\frac{t(2-\alpha+\alpha e^{2ika})}{2(1-e^{2ikL-+})}
    %\end{equation}
    %\begin{equation} \label{45}
    %F_1=\frac{-te^{2ikL_-}(2-\alpha+\alpha e^{2ika})}{2(1-e^{2ikL_-})}
    %\end{equation}
    %\begin{equation} \label{46}
    %C_2=\frac{te^{ika}}{1-e^{2ikL_+}}
    %\end{equation}
    %\begin{equation} \label{47}
    %D_2=\frac{-te^{2ikL^+}e^{ika}}{1-e^{2ikL_+}}
    %\end{equation}
    %\begin{equation} \label{48}
    %G_2=\frac{te^{ika}}{1-e^{2ikL_-}}
    %\end{equation}
    %\begin{equation} \label{49}
    %D_2=\frac{-te^{2ikL^-}e^{ika}}{1-e^{2ikL_-}}
    %\end{equation}

    \begin{equation} \label{50}
        \begin{split}
            D_m  = \int_{x=0}^{x=a} [\frac{2+\abs{\alpha}^2}{2}+\frac{-\abs{\alpha}^2-2\alpha}{4}e^{2ik(x-a)}+\\
            \frac{-\abs{\alpha}^2+2\alpha}{4}e^{-2ik(x-a)}]\abs{t}^2 dx
        \end{split}
    \end{equation}
For the vertical region of the left crossbar we have:
    \begin{equation} \label{51}
        \begin{split}
            D_1 & =\int_{y=-L^-}^{y=L^+} \abs{\Psi_1}^2dy\\
            & =\int_{y=0}^{y=L^+} \abs{\Psi_1^A}^2dy + \int_{y=-L^-}^{y=0} \abs{\Psi_1^B}^2dy
        \end{split}
    \end{equation}
With $\Psi_1^A$ and $\Psi_1^B$ the wave functions for the upper and lower arm respectively, and being their quadratic values:
    \begin{equation} 
    \begin{split}
    \label{52}
        \abs{\Psi_1^A}^2=\frac{\abs{t}^2[2-2cos(2k(y-L^+))]}{2[2-2cos(2kL^+)]}\times\\
        \times[2-2\abs{\alpha}sin(2ka)+\abs{\alpha}^2(1-cos(2ka))]
    \end{split}
    \end{equation}
    \begin{equation} 
    \label{53}
    \begin{split}
    \abs{\Psi_1^B}^2=\frac{\abs{t}^2[2-2cos(2k(y-L^-))]}{2[2-2cos(2kL^-)]}\times\\
        \times[2-2\abs{\alpha}sin(2ka)+\abs{\alpha}^2(1-cos(2ka))]
    \end{split}
    \end{equation} 
And for the vertical region of the right crossbar we have:
    \begin{equation} \label{54}
        \begin{split}
            D_2 & =\int_{y=-L^-}^{y=L^+} \abs{\Psi_2}^2dy\\
            & =\int_{y=0}^{y=L^+} \abs{\Psi_2^A}^2dy + \int_{y=-L^-}^{y=0} \abs{\Psi_2^B}^2dy
        \end{split}
    \end{equation} 
With $\Psi_2^A$ and $\Psi_2^B$ the wave functions for the upper and lower arm respectively, and being their quadratic values:
    \begin{equation} \label{55}
        \abs{\Psi_2^A}^2=\frac{\abs{t}^2[1-cos(2k(y-L^+))]}{1-cos(2kL^+)}
    \end{equation}
    \begin{equation} \label{56}
        \abs{\Psi_1^B}^2=\frac{\abs{t}^2[1-cos(2k(y-L^-))]}{1-cos(2kL^-)}
    \end{equation}

\bibliography{biblio-bic}

\end{document}